# Optimal Cooling of Multiple Levitated Particles through Far-Field Wavefront-Shaping

Jakob Hüpfl[1,†], Nicolas Bachelard[1,2,†,*], Markus Kaczvinszki[1], Michael Horodynski[1], Matthias Kühmayer[1], and Stefan Rotter[1,*]

**Affiliations:**

[1] Institute for Theoretical Physics, Vienna University of Technology (TU Wien), A–1040 Vienna, Austria

[2] CNRS, LOMA, UMR 5798, Université de Bordeaux, Talence, France

[†] These authors contributed equally to this work.

[*] To whom correspondence should be addressed: nicolas.bachelard@gmail.com, stefan.rotter@tuwien.ac.at

**Abstract:**

Light forces can be harnessed to levitate mesoscopic objects and cool them down towards their motional quantum ground state. Significant roadblocks on the way to scale up levitation from a single to multiple particles in close proximity are the requirements to constantly monitor the particles' positions as well as to engineer complex light fields that react fast and appropriately to their movements. Here, we present an approach that solves both problems at once. By exploiting the information stored in a time-dependent scattering matrix, we introduce a robust formalism enabling the identification of spatially modulated wavefronts, which simultaneously cool down multiple levitated objects of arbitrary shapes. An experimental implementation is suggested based on stroboscopic scattering-matrix measurements and time-adaptive injections of modulated light fields.

**Main Text:**

The desire to exploit light for the manipulation of matter has led to remarkable achievements such as optical tweezers [1], the laser-cooling of gases [2] or the realization of Bose-Einstein condensates [3,4]. A recent and exciting endeavor lies in using laser light to cool mesoscopic objects down to their motional quantum ground state [5]. To further decouple these objects from their environment and to make them directly accessible through optical micro-manipulation, one laser-levitates them in vacuum [6]. While promising remarkable opportunities for high-resolution sensing [7–9], or for testing the limits of quantum physics [10,11], levitation so far strongly relies on the accessibility of local information. Take, as example, tweezer-assisted cavity-cooling schemes that were recently applied to reach the ground state of a nanometer-size bead [12] through an approach known as coherent scattering [13–15]. There, performances are constrained by the ability to accurately position the object within the optical mode [16–18]. In feedback-cooling schemes [19], the position of a trapped particle needs to be constantly monitored to bring the system to its ground state [20,21]. Yet, such monitoring leads to major calibration issues that ultimately affect the efficiency of these techniques [22]. Alongside the difficulty of multiplexing traps in close proximity due to optical binding [23], these limitations prevent levitation from being scaled-up to multiple coupled particles or to be applied simultaneously to different motional degrees of freedom.

Yet, even when light from the control laser gets scattered by one or multiple levitated objects, it carries information about the objects' geometry and motion towards the far field. The book-keeping of this information is conveniently organized in the scattering matrix, which connects the spatial profiles (i.e., wavefronts) of incoming and outgoing scattering states. Routinely measured even for very complex systems [24–26], the scattering matrix has already provided access to tailor-made light fields for applications in imaging [27–30], opto-mechanics [31–33] or quantum optics [34,35].

In this Letter, we demonstrate a novel and straightforward procedure to distill from the scattered far field the wavefronts necessary for the manipulation of several levitated objects in parallel. This approach can cool down or heat up multiple particles of non-trivial shapes experiencing complex motion. Notably, our cooling scheme also applies to multiple coupled opto-mechanical resonators that are realized by nano-objects trapped at different maxima of a standing wave. With its capability to handle different motional degrees of freedom simultaneously, our procedure also turns out to be remarkably robust to limitations in the availability of scattered-field information as necessary to be compatible with state-of-the-art levitation setups.

Our starting point is the measurable scattering matrix $\boldsymbol{S}$, which relates any incoming wave on a medium, $|\Psi_{\mathrm{in}}\rangle$, to the outgoing field that is scattered towards the far field, $|\Psi_{\mathrm{out}}\rangle = \boldsymbol{S}|\Psi_{\mathrm{in}}\rangle$ (see Fig. 1(a)). To get direct access to the observable of interest, $\boldsymbol{S}$ must be recast into a different linear operator that represents this observable. Take as an example, here, a static scattering system, where the information on the time involved in the scattering process is represented by the time-delay (TD) operator, $\boldsymbol{Q}_{\mathrm{TD}} = -i\boldsymbol{S}^{\dagger}\partial_{\omega}\boldsymbol{S}$, introduced by Wigner and Smith [36,37]. Featuring a derivative with respect to the laser light's angular frequency $\omega$, this Hermitian operator $\boldsymbol{Q}_{\mathrm{TD}}$ contains the time that each of its eigenstates spends inside the scattering region as a corresponding real eigenvalue. With regard to our goal to cool down an ensemble of moving particles, however, the observable of interest is not the time delay of scattering states, but the shift in the levitated particles' total energy induced by the incoming light field. Moreover, rather than being static, the system we consider here follows a dynamic yet slow evolution, i.e., on a time-scale larger than the typical time delay of

the scattered light fields. As we show in detail in an accompanying article [38], the linear scattering operator that provides access to this energy shift (ES) turns out to be a different Wigner-Smith operator, $\boldsymbol{Q}_{\mathrm{ES}}(t) = -i\boldsymbol{S}^{\dagger}(t)\partial_t \boldsymbol{S}(t)$, involving a time-derivative $\partial_t$ of the instantaneous scattering matrix $\boldsymbol{S}(t)$ that is dynamically changing due to the particles' motion and measured in the far field at time $t$. A variant of this Hermitian operator was introduced by Avron *et al*. (with $\boldsymbol{S}^{\dagger}$ and $\partial_t \boldsymbol{S}$ interchanged) in the context of mesoscopic electron transport [39] to describe how externally driven charge pumps pass electrons through a conductor [40]. Here, we study the reverse situation: rather than operating a fermionic charge pump by a temporal change of the scattering system, we inject a suitably shaped bosonic light field to induce an opto-mechanical modification of the system itself. Importantly, injecting eigenstates of the energy-shift operator $\boldsymbol{Q}_{\mathrm{ES}}(t)$ changes a collective property of the system (its total energy) rather than just the motion of individual constituents [32].

At any given time, the energy-shift operator can be harnessed to identify wavefronts that generate optical forces able to instantaneously reduce the total mechanical energy of a multi-particle system. Figure 1(a) displays a corresponding setup composed of a multimode waveguide filled with nano-objects of arbitrary shapes that experience random motions. Two spatial light modulators (SLMs) distributed on both sides of the system serve to constantly measure the instantaneous scattering matrix, $\boldsymbol{S}(t)$, and to inject spatially modulated wavefronts, $|\Psi_{in}(t)\rangle$. In the waveguide, any individual object of mass $m$ and speed $\vec{v}_n$ executes an underdamped motion that fulfills

$$m\frac{d\vec{v}_n(t)}{dt} = \vec{F}_n(t) - m\gamma\vec{v}_n(t) + m\vec{g} + \vec{\xi}(t), \tag{1}$$

in which $\vec{F}_n$ stands for the optical force produced by $|\Psi_{in}\rangle$, $\gamma$ for the friction coefficient of the environment, $\vec{g}$ for gravity and $\vec{\xi}(t)$ for a white-noise process describing the coupling to the surrounding thermal bath. When the power of non-conservative forces (e.g., friction) remain smaller than the variations in total mechanical energy of the system, $E_{tot}$, we demonstrate in [38] that an incoming field, $|\Psi_{in}\rangle$, produces optical forces that shift the total energy of the multi-particle system by an amount quantified by the energy-shift operator

$$\langle\Psi_{in}|\boldsymbol{Q}_{\mathrm{ES}}(t)|\Psi_{in}\rangle = \frac{4\pi c}{\lambda}\frac{dE_{tot}(t)}{dt}, \tag{2}$$

where $\lambda$ stands for the optical wavelength and $c$ the speed of light. From Eq. (2), we readily deduce that, at time $t$, the real eigenstate $|\Psi_{min}(t)\rangle$ of the Hermitian operator $\boldsymbol{Q}_{ES}(t)$ corresponding to its minimal (i.e., most negative) eigenvalue, $\theta_{min}$, will perform an optimal reduction of system's energy ($d_t E_{tot}(t) = \theta_{min}\lambda/4\pi c$). We will thus refer to $|\Psi_{min}(t)\rangle$ as the optimal cooling state. For the same reason, optimal heating (i.e., increase of $E_{tot}$) will be performed by the eigenstate of $\boldsymbol{Q}_{ES}(t)$ with the highest eigenvalue. Importantly, $E_{tot}$ encompasses here both translational as well as rotational degrees of freedom and the states $|\Psi_{min}(t)\rangle$ are readily extracted from $\boldsymbol{Q}_{ES}(t)$. The energy-shift matrix $\boldsymbol{Q}_{ES}(t)$ itself is determined only by far-field measurements and comes without any prior knowledge of the particles' geometry and motion. Introduced here for a waveguide setup, Eq. (2) remains valid also for objects in free space that experience complex three-dimensional motion [41].

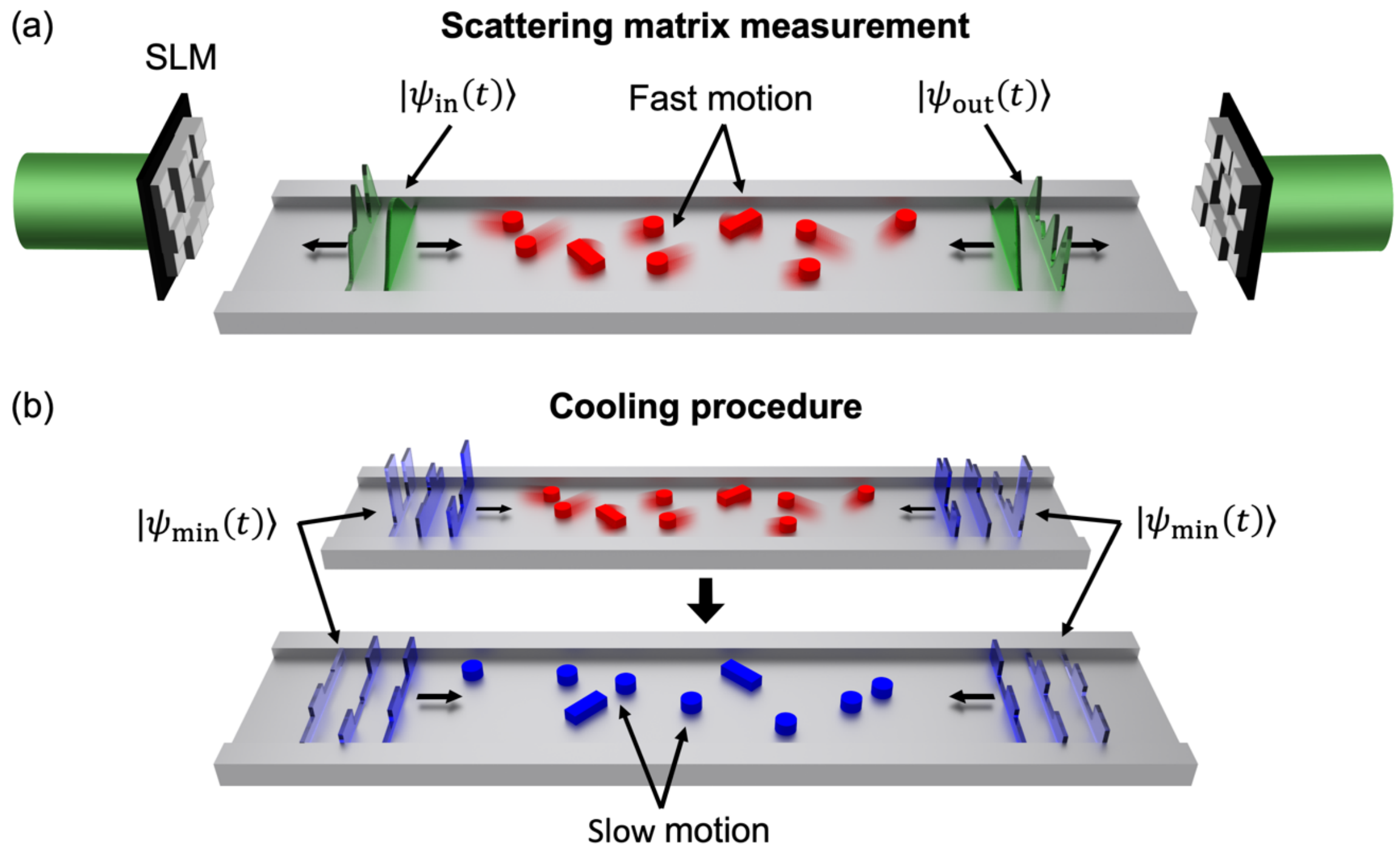

**Fig. 1 |** (a) A multimode waveguide (grey) is filled with dielectric moving nano-objects with different shapes (bars and cylinders). SLMs are placed on both sides of the waveguide to inject spatially modulated fields $|\Psi_{\text{in}}(t)\rangle$ onto the system. The fields that are scattered out, $|\Psi_{\text{out}}(t)\rangle$, are recorded and serve to measure the instantaneous scattering matrix, $\boldsymbol{S}(t)$, that fulfills $|\Psi_{\text{out}}(t)\rangle = \boldsymbol{S}(t)|\Psi_{\text{in}}(t)\rangle$. (b) Initially, the gas of nano-particles follows a random motion, in which individual objects rotate and/or translate (red bars and beads). A succession of optimal cooling wavefronts, $|\Psi_{\text{min}}(t)\rangle$, gets injected from both leads (blue wavefronts) to optimally slow the particles' motion at respective times $t$, which ultimately cools down the gas (blue bars and beads).

As illustrated in Fig. 1(b), for an underdamped gas of randomly moving objects (red particles), applying a succession of optimal cooling states at sampled time steps ($|\Psi_{min}(t)\rangle$) effectively produces an artificial damping that cools down the ensemble collectively (blue particles). In Fig. 2(a), we present results from the numerical simulation of an ensemble of $N = 10$ silica beads (radius $r = 75$ nm, refractive index $n = 1.44$) that displays a low friction motion in the $(x, z)$ plane fulfilling Eq. (1). The gas is initially thermalized and its velocities follow a Boltzmann distribution of mean absolute value $\bar{v}_0 = 13\ \mathrm{mm/s}$ (corresponding to an initial temperature of $30\ \mathrm{K}$). Here, the influence of gravity in Eq. (1) is negligible and the total energy results solely from its kinetic contribution. The particles are confined within a multimode waveguide featuring $M = 10$ transverse modes and they experience optical forces produced by a monochromatic field (wavelength $\lambda = 532\ \mathrm{nm}$). $\boldsymbol{S}(t)$ is measured at a sampling rate $\Delta t_{cool} = 1\ \mu\mathrm{s}$ and the corresponding energy-shift operator is approximated by $\boldsymbol{Q}_{ES}(t) \approx -i\boldsymbol{S}^{\dagger}(t)[\boldsymbol{S}(t) - \boldsymbol{S}(t - \Delta t_{cool})]/\Delta t_{cool}$. The initial configuration of the particles is shown in Fig. 2(a) at $t = 0$ μs, together with the optimal cooling state $|\Psi_{min}(t = 0)\rangle$ extracted from $\boldsymbol{Q}_{\mathrm{ES}}(t = 0)$. This state gets injected with an optical power $P_{\mathrm{in}} = 20\ nW$ for $\Delta t_{\mathrm{cool}} = 1$ μs, before the new optimal cooling state (corresponding to the new $\boldsymbol{Q}_{\mathrm{ES}}(t + \Delta t_{\mathrm{cool}})$) is computed and injected for the same duration. The process is then iteratively repeated every $\Delta t_{\mathrm{cool}}$ (see Fig. 2(a) for snapshots of the particles and the injection fields at $t = 50$ and $100$ μs) with Fig. 2(b) displaying the corresponding time evolution of the total energy. We observe that the energy continuously drops by more than three orders of magnitude in about $100$ μs, thus corroborating that successively applying optimal wavefronts assembled from Eq. (2) acts as an “artificial” damping. In analogy to single-element feedback cooling [19], this damping is non-dissipative and results in unconventional entropy production [42]. For comparison, Fig. 2(c) shows the anticipated mechanical action on the gas when an unmodulated wavefront (the fundamental transverse mode) gets injected into the waveguide with the same power ($P_{\mathrm{in}} = 20$ nW). This field randomly transfers momentum to individual elements such that the energy of the gas remains almost unchanged for short timescales (Fig. 2(d)). For longer timescales (i.e., comparable with the dissipation rate), the system heats up until a macroscopic thermal steady state is reached (not shown). In [38], we show that the cooling efficiency is limited by the sampling time (shorter $\Delta t_{\mathrm{cool}}$ provides a prompter response to the particles' movements). Moreover, the cooling performance is maximized for a specific optical power that optimally balances the particles' motion. Videos showing the time evolution of different particle ensembles cooled through the successive injections of optimal-cooling states are provided in the Supplementary Movies M1 and M2.

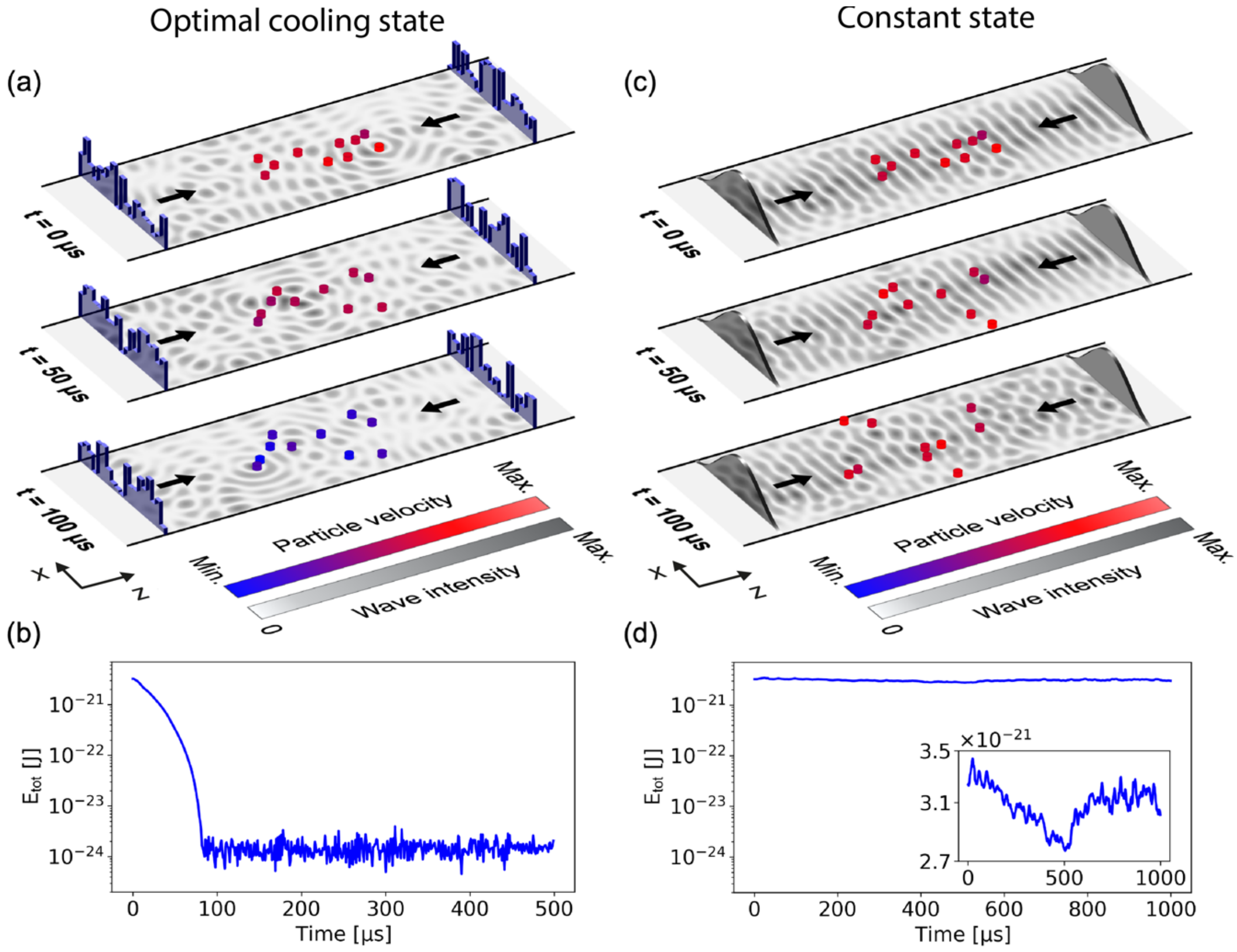

**Fig. 2 |** (a) Confined in a waveguide with $M = 10$ transverse modes, a gas of $N = 10$ nanometer-size spherical particles ($r = 75\ nm, m = 4.7 \times 10^{-18}\ kg, \gamma = 6\ Hz$) is initially in random motion (red beads in top panel at $t = 0\ \mu s$). Over time, the gas is submitted to a succession of optimal cooling wavefronts (blue discontinuous lines on both leads) producing complex scattered fields in the waveguide (black and white intensity). At each time step in the numerical simulation, the $\boldsymbol{S}$-matrix corresponding to the current location of the particles is evaluated, $\boldsymbol{Q}_{\mathrm{ES}}(t)$ is computed and its lowest eigenstate $|\Psi_{\min}(t)\rangle$ is applied to reduce the total energy of the gas. While these optimal cooling states are successively applied, the speeds of individual particles are shown to decrease progressively, as indicated by their colors in the three panels for times $t = 0\ \mu s$ (top), $50\ \mu s$ (middle) and $100\ \mu s$ (bottom), which gradually

transition from red to blue. (b) Log-scale evolution of the total energy, $E_{\mathrm{tot}}$, of the gas during the procedure of (a). (c) For comparison, the gas shown in (a) is submitted to a constant incoming wavefront corresponding to the fundamental transverse mode of the waveguide (grey curves in both leads). In contrast to (a), the particles can be observed to wander around without cooling down. (d) Log-scale evolution of the total energy, $E_{\mathrm{tot}}$, of the gas during the procedure of (c). The small fluctuations in energy are shown in the zoom-in plot in the inset.

The energy-shift operator can also serve to simultaneously cool coupled resonators consisting of multiple trapped nano-objects. We illustrate this in Fig. 3(a), where a trapping laser field (green shape, $|\Psi_{trap}\rangle$, $\lambda_{trap} = 1550\,nm$) gets injected from both sides in the waveguide's fundamental transverse mode to form a standing wave, whose maxima correspond to potential wells (green concentric contours). Five silica nano-beads ($r = 75\,nm$) are confined within these wells and form a chain of coupled opto-mechanical resonators. Along the longitudinal direction $z$, each oscillator is characterized by a power spectral density, $|S_{zz}|$, displaying a main individual resonance close to 40 kHz, whose frequency varies slightly depending on a particle's position along the chain, while the coupling amongst particles manifests itself through the presence of harmonics (blue curve, Fig. 3(c)). Using a second laser, we now apply the same iterative procedure as described in Figs. 2(a) and (b) by sending a succession of optimal cooling states, $|\Psi_{min}(t)\rangle$, while the trapping field remains in place. The corresponding dynamics of individual particles is then described by Eq. (1) in which a trapping-force contribution is added, while, according to Eq. (2), the cooling procedure reduces $E_{tot}$ that now encompasses both the kinetic and the potential energies of all the trapped particles. As with conventional single-object cooling [19], by reducing the total energy of the resonators, our procedure pins the different nano-objects at the bottom of their individual potentials. Using a power of $200\,nW$, Fig. 3(b) displays the time evolution of the combined fluctuations along $z$ of all the particles around their respective trapping positions $z_n$, which decrease by several orders of magnitude throughout the cooling process. As expected, the procedure effectively increases motional damping and ultimately broadens individual resonances. Figure 3(c) displays in blue the power spectral density, $|S_{zz}|$, of the leftmost particle along the chain before cooling. For the orange and green spectra, the procedure is then performed with $20$ and $200\ nW$ to cool the system down to respective center-of-mass temperatures of $25$ and $0.65\ mK$ (temperatures extracted through a fit of $|S_{zz}|$ [43]). Remarkably, when the particles get cooled, we observe that the coupling-induced harmonics disappear, while the quality factor of their main resonance reduces with cooling power until reaching a minimum for roughly $200\ nW$ (similar behavior observed for all trapped nano-beads). A video showing the cooling of the five particles in Fig. 3 is provided in Supplementary Movie M3.

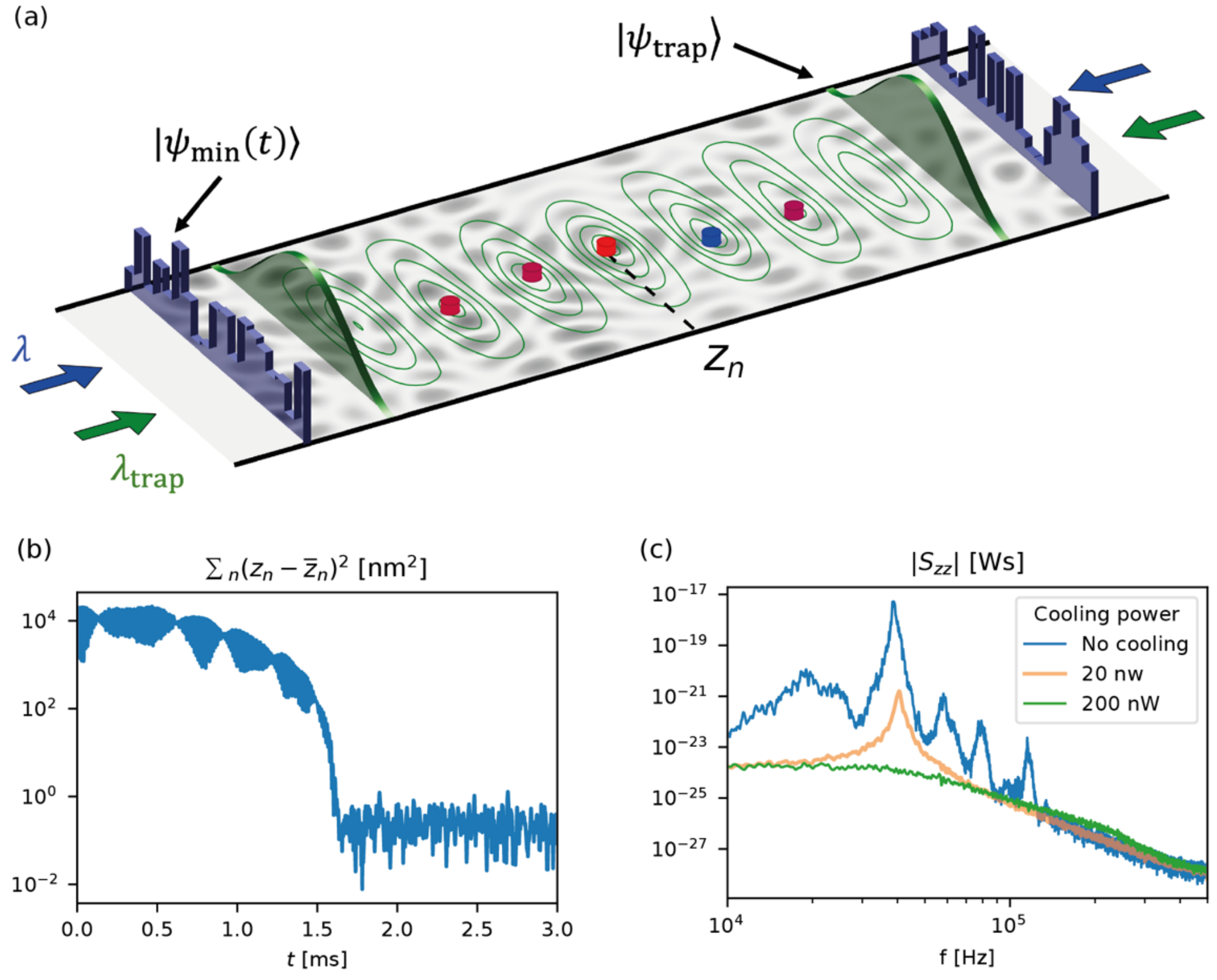

**Fig. 3 |** (a) A trapping laser field (green waves, $|\Psi_{trap}\rangle$) gets injected on both sides of the waveguide to form a standing wave (green concentric contours). Five nano-beads (positions $z_{n\in[1,5]}$) are confined at the middle of five antinodes of the standing wave (mean positions $\bar{z}_{n\in[1,5]}$). The cooling procedure of Figs. 2(a) and (b) is applied by injecting a succession of optimal cooling states, $|\Psi_{min}(t)\rangle$ (blue wavefronts) that reduce the system's total energy. (b) Log-scale time evolution of the variance of the particles' positions around their individual trapping locations (i.e., $\sum_n (z_n - \bar{z}_n)^2$) throughout the cooling procedure and using an optical power of $200\, nW$. (c) Within its trap, each particle forms an opto-mechanical resonator. The blue curve displays the power spectral density, $|S_{zz}|$, of the resonator at $z_1$ before cooling, while the orange and green curves correspond to cooling performed at $20$ and $200$ nW, respectively.

While our derivations implicitly rely on the assumption that the scattering matrix $\boldsymbol{S}(t)$ is unitary (i.e., loss-free), the implementation of optimal cooling states turns out to be robust to missing information, such as when the scattering matrix is incomplete or the particles are absorbing. For a gas made of $10$ particles confined in a waveguide with $M = 20$ transverse modes, Fig. 4(a) shows the performance of the protocol introduced in Figs. 2(a) and (b) with $\boldsymbol{Q}_{\mathrm{ES}}(t)$ now being assembled from an incomplete set of modes, i.e., $\boldsymbol{S}(t)$ is expressed within the basis of the first $M_P \leq 20$ modes. The green, orange and blue curves correspond to the temporal evolution of $E_{\mathrm{tot}}(t)$ obtained for $M_p = 8,\ 4$ and $2$, respectively, and all of them are observed to lead to significant cooling. For comparison, the black curve reports the cooling performed with the full set of modes (i.e., $M_p = M$). Figure 4(b) considers a complete set of $M = 20$ modes while absorption is now introduced within the particles through an imaginary part, $n_i$, of their refractive index. The orange, green and red curves display the cooling reported for $n_i = 0.01,\ 0.05$ and $0.1$, respectively. For comparison, the black curve reports the cooling performed without absorption (i.e., $n_i = 0$). Here, we do not take into account measurement imprecisions that typically impact cooling schemes [44–46]. Yet, together, Figs. 4(a) and (b) suggest that our approach can operate under partial information collection and degraded $\boldsymbol{S}$-matrix reconstruction. We thus anticipate that this scheme should also be able to handle measurement imprecisions. Nonetheless, when reducing the available information, the performance of our cooling scheme degrades: the convergence time increases when $M_p$ decreases or when $n_{\mathrm{i}}$ increases, while the cooling efficiency worsens when decreasing $M_p$ in Fig. 4(a) and, ultimately, fails for $n_i \geq 0.6$ in Fig. 4(b). At last, we emphasize here that we neglected heating by laser phase noise and other mechanisms that are known to be problematic in high-vacuum conditions [12,16].

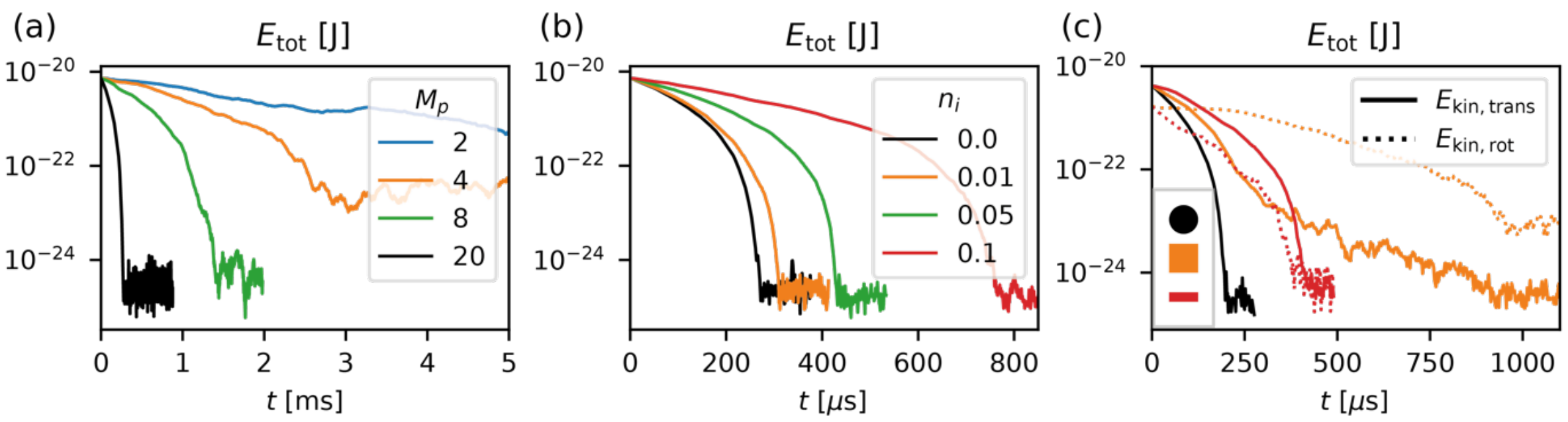

**Fig. 4 |** (a), Total energy over time in log-scale for a gas of $N = 10$ particles (without trapping field), cooled using $M_p = 20$ (black), $8$ (green), $4$ (orange) and 2 (blue) modes amongst the $M = 20$ transverse modes in the waveguide, respectively. Less information (i.e., smaller $M_p$) results in longer cooling times and reduces cooling performances. (b) Total energy over time in log-scale for a gas of $N = 10$ particles, cooled in a $M = 20$ multimode waveguide considering an intrinsic absorption of the particles that is parameterized by the imaginary part of their refractive index, $n_i$. The black, orange, green and red curves correspond to an absorption of $n_i = 0,\ 0.01,\ 0.05$ and $0.1$, respectively. Stronger absorption results in longer cooling times. (c) Log-scale evolution of the total energy associated with translational (continuous curves) and rotational (dashed curves) degrees of freedom during the cooling of gases composed of $N = 10$ squares (sidelength $210\ nm$, orange curves) and $N = 10$ rectangles ($290 \times 73\ nm$, red curves). At $t = 0$, the translational energies appear twice larger than the rotational ones as the particles experience two degrees of freedom in translation compared to one in rotation. For comparison, the black curve shows the cooling of a $N = 10$ nano-beads (radius $150\ nm$).

We emphasize that our procedure remains effective for non-trivial particle shapes. Figure 4(c) displays in orange (red) the cooling of a gas of $N = 10$ squares (rectangles) that are initially in a random combination of translation and rotation. For comparison, the black curve shows the cooling of a gas composed of nano-beads. The solid and dashed curves represent the evolution of the kinetic energy related to translational and rotational motions, respectively, which are systematically reduced by the cooling procedure. Yet, we observe that rotational degrees of freedom are easier to cool down for rectangles as compared to squares as their geometry provides more pronounced edges that enable to efficiently exert optical torques. The Supplementary Movie M4 shows that an inhomogeneous mixture of differently shaped particles can also be cooled down robustly.

Finally, we stress that our procedure can be implemented with current state-of-the-art modulator technology. As explained in [38], the maximum stroboscopic timespan still allowed for experimentally cooling a dilute gas of sub-wavelength particles can be estimated by $\Delta t_{\mathrm{max}} \approx \lambda/4\bar{v}_0$, i.e., the ratio of the optical cooling wavelength, $\lambda$, over the particles' root-mean-square velocity, $\bar{v}_0$. For the systems considered here, this timespan lies around $\Delta t_{\mathrm{max}} \approx 10$ µs, which is consistent with the performance of current spatial light modulators that can operate way above the MHz range [44,45] and are expected to reach the GHz range soon [49].

In summary, we use scattered-field information to capture the collective motion of a complex system composed of mesoscopic objects. Assembled from the scattering matrix, a linear energy-shift operator enables the simultaneous manipulation of multiple motional degrees of freedom to perform the cooling or heating of such many-body systems. Implementable with current optical modulators, our approach is robust against information losses and neither requires the detection of particles nor calibration. With its flexibility with respect to the particles' dimensions or shapes, our method could prove to be a crucial tool for quantum-state engineering in mesoscopic many-body systems [50]. By providing access to macroscopic system properties through the scattering matrix, our work also opens up new directions in non-equilibrium thermodynamics [51], such as for the realization of complex nano-heat engines [52] or for the assembly of dynamical materials [53,54].

Acknowledgments:

We thank J. Bertolotti and Y. Louyer for helpful discussions and the team behind the open-source code NGSolve for assistance. Support by the Austrian Science Fund (FWF) under Project No. P32300 (WAVELAND) and funding for NB from the European Union's Horizon 2020 research and innovation program under the Marie Skłodowska-Curie grant agreement No. 840745 (ONTOP) are gratefully acknowledged. The computational results presented were achieved using the Vienna Scientific Cluster (VSC).

Jakob Hüpfl and Nicolas Bachelard contributed equally to this work.